\documentclass[fleqn,usenatbib]{mnras}
\pdfoutput=1 
\usepackage{newtxtext,newtxmath}

\usepackage[T1]{fontenc}

\usepackage{graphicx}	

\newcommand{\hmpc}{h^{-1}{\rm Mpc}}
\newcommand{\hgpc}{h^{-1}{\rm Gpc}}

\newcommand{\mbr}{\boldsymbol r}
\newcommand{\cdf}{{\rm CDF}}

\newcommand{\nn}{{\rm NN}}

\newcommand{\eq}[2]{\begin{align} \label{eq:#1} #2 \end{align}}
\newcommand{\ns}{\!\!}
\newcommand{\mathp}{\mathcal{P}}
\newcommand{\dd}{{\rm d}}
\newcommand{\quijote}{\textsc{Quijote}\,}
\newcommand{\corrfunc}{\textsc{Corrfunc}\,}

\title[Cross-correlations and NN Distributions]{Cosmological cross-correlations and nearest neighbor distributions}

\author[Banerjee \& Abel]{
Arka Banerjee $^{1,2,3,4}$\thanks{E-mail: {\tt arkab@stanford.edu}}
and Tom Abel $^{2,3,4}$\thanks{E-mail: {\tt tabel@stanford.edu}} \\
$^{1}$Fermi National Accelerator Laboratory, Cosmic Physics Center, Batavia, IL 60510, USA \\
$^{2}$Kavli Institute for Particle Astrophysics and Cosmology, Stanford University, 452 Lomita Mall, Stanford, CA 94305, USA \\
$^{3}$Department of Physics, Stanford University, 382 Via Pueblo Mall, Stanford, CA 94305, USA \\
$^{4}$SLAC National Accelerator Laboratory, 2575 Sand Hill Road, Menlo Park, CA  94025, USA
}

\date{Accepted XXX. Received YYY; in original form ZZZ}

\pubyear{2021}
\begin{document}
\label{firstpage}
\pagerange{\pageref{firstpage}--\pageref{lastpage}}
\maketitle

\begin{abstract}
Cross-correlations between datasets are used in many different contexts in cosmological analyses. Recently, $k$-Nearest Neighbor Cumulative Distribution Functions ($k\nn$-$\cdf$) were shown to be sensitive probes of cosmological (auto) clustering. In this paper, we extend the framework of nearest neighbor measurements to describe joint distributions of, and correlations between, two datasets. We describe the measurement of \textit{joint} $k\nn$-$\cdf$s, and show that these measurements are sensitive to all possible connected $N$-point functions that can be defined in terms of the two datasets. We describe how the cross-correlations can be isolated by combining measurements of the joint $k\nn$-$\cdf$s and those measured from individual datasets. We demonstrate the application of these measurements in the context of Gaussian density fields, as well as for fully nonlinear cosmological datasets. Using a Fisher analysis, we show that measurements of the halo-matter cross-correlations, as measured through nearest neighbor measurements are more sensitive to the underlying cosmological parameters, compared to traditional two-point cross-correlation measurements over the same range of scales. Finally, we demonstrate how the nearest neighbor cross-correlations can robustly detect cross correlations between sparse samples --- the same regime where the two-point cross-correlation measurements are dominated by noise.
\end{abstract}

\begin{keywords}
cosmology 
\end{keywords}



\section{Introduction}
\label{sec:intro}

Measurements of statistical cross-correlations between datasets are widely used in cosmology and astrophysics, often in the context of spatial clustering of the data. These measurements, which characterize the spatial correlations in the fluctuations in the density field or number counts of the two datasets, have a number of uses, depending on the context  \citep[see e.g.][]{2013arXiv1309.5388R}. For example, cross-correlations can help break various degeneracies and allow for stronger constraints on parameters of interest. The measurement of galaxy-galaxy lensing - the cross-correlation of galaxy counts and weak lensing maps is widely used, in conjunction with measurements of galaxy clustering and cosmic shear, to break degeneracy between halo bias and cosmological parameters \citep[e.g.][]{2005PhRvD..71d3511S, 2013MNRAS.432.1544M, 2015ApJ...806....1M, 2015ApJ...806....2M, 2018PhRvD..98d3526A, 2018MNRAS.474.4894J, 2020MNRAS.492.2872W, 2020arXiv200715632H}. Similarly, cross-correlations between galaxy populations and galaxy clusters can also break individual bias degeneracies, and better constrain cosmological parameters \citep[e.g.][]{1999MNRAS.305..547C, 2013MNRAS.431.3319Z, 2017MNRAS.470.2566P, 2020MNRAS.491.3061S, 2020arXiv201001138T}. In other contexts, cross-correlations can be used to mitigate the effects of survey systematics on the cosmology analysis. This is especially relevant for datasets which are measured in different surveys - while the cosmological signal is correlated, the systematics between the two surveys are usually uncorrelated. Examples of this aspect is the use of cross-correlations between Cosmic Microwave Background (CMB) lensing maps, and galaxy-galaxy lensing and galaxy clustering \citep[e.g.][]{2016MNRAS.459...21K, 2016MNRAS.461.4099B, 2017MNRAS.464.2120S, 2017PhRvD..95l3512S, 2019PhRvD.100b3541A,  2020MNRAS.491...51S}. 

Cross-correlations and their applications in cosmology can be roughly divided into two categories, based on the statistical significance of the relevant signal,. In the high signal-to-noise regime, the variations in the cross-correlation signal as a response to a change in the underlying cosmological parameters are large compared to the errors in the measurement process. Therefore, the cross-correlation measurements can be used to infer the values of the cosmological parameters. This is precisely how the cross-correlation measurements are used in most of the references cited above. On the other hand, when the signal-to-noise ratio is low, the focus is primarily on the detection of a cross-correlation signal, rather than its use directly in parameter inference \citep[e.g.][]{2006MNRAS.368..732B, 2008ApJ...683L..99G,2008arXiv0805.2974G, 2015ApJ...802...64B, 2019arXiv190210120L, 2019ApJ...882...62N, 2020PhRvL.124j1102A, 2020ApJ...894..112F}. A common theme in this regime is to look for cross-correlations between rare but interesting astrophysical signals and a set of relatively dense, well-calibrated tracers of Large Scale Structure (LSS). The rare events usually have such low number densities that their auto-clustering is completely dominated by noise, but the clustering signal can be recovered through the cross-correlations, which is not affected by shot noise. This technique has been explored in the context of cross-correlating ultra high-energy neutrino sources detected by IceCube with galaxies \citep{2020ApJ...894..112F}; gamma ray sources with weak lensing measurements \citep{2020PhRvL.124j1102A}; and Fast Radio Bursts (FRBs) with galaxies \citep{2019arXiv190210120L}.

It is worth noting that the term ``cross-correlations'' in the context of cosmology generally refers to the two-point cross-correlations of the datasets. The two-point cross correlations capture the full information between two Gaussian fields. However, at late times, and on small scales, cosmological density fields can be highly nonlinear, and depart strongly from a Gaussian distribution. Therefore, correlations between two of these fields can exist beyond the two-point cross-correlation. These higher order cross-correlations can, in principle, be used to better characterize clustering of the two fields \citep[e.g.][]{2005A&A...432..783S, 2014MNRAS.442...69M, 2017A&A...606A.128R}. 

Recently, \citet{Banerjee_Abel} introduced a new approach to studying clustering in the cosmological data --- through the use of $k$-Nearest Neighbor Cumulative Distribution Functions ($k\nn$-$\cdf$). This is the empirical cumulative distribution function of distances from a set of
volume-filling, Poisson distributed random points to the $k$–nearest data points, and has various attractive properties. It is computationally inexpensive to measure, and is formally sensitive to all connected $N$-point functions of the continuous field from which the data points are drawn. \citet{Banerjee_Abel} demonstrated that these summary statistics are more sensitive to the underlying cosmological parameters than the two point auto-correlation function (over the same range of scales), and therefore, promises to be a useful tool to optimally extract information from small scales in cosmological surveys.

Whereas \citet{Banerjee_Abel} focused on the clustering of only one set of tracers, in this paper, we extend the theoretical and measurement framework of $k\nn$ distributions to describe the joint clustering of two sets of tracers, enabling the same formalism to describe both auto-correlations and cross-correlations. We demonstrate how the \textit{joint} nearest neighbor distributions are formally sensitive to all $N$-point functions that can be defined by the two fields, and identify the parts that are sensitive to the cross-correlations. In this context, the term  ``cross-correlations" refer to any statistical dependence of the two fields with each other, not just the traditional two-point correlations. We outline how to measure these joint distributions efficiently using distances to nearest neighbor data points of each set from a set of dense volume-filling randoms, as well as the method to measure only the cross-correlation piece from the same set of measurements. We apply these measurements in the context of two Gaussian tracers, where the distributions can be predicted analytically, and then to tracers of fully nonlinear fields. For the latter, we demonstrate the statistical power of the $k\nn$ cross-correlations compared to the two-point cross-correlations through two examples - one in the high signal-to-noise regime and one in the low signal-to-noise regime. Given the fact that cross-correlations are used so widely in cosmological analyses, this extension of the $k\nn$ formalism and measurements vastly increases the range of possible applications. 

The paper is arranged as follows: in Sec. \ref{sec:NNCDF}, we introduce the framework of joint nearest neighbor distributions for two correlated fields, and outline how these are measured on actual data. Then, in Sec. \ref{sec:Gaussian}, we consider the case of joint $k\nn$ distributions and cross correlations for tracers of correlated Gaussian fields. In Sec. \ref{sec:LSS}, we apply the same measurements to sets of tracers of fully nonlinear fields. We compare the sensitivity of $k\nn$ cross-correlation measurements  of simulation halos and matter field to underlying cosmological parameters, and compare it to that of two-point cross-correlations. We also demonstrate the detection of a cross-correlation signal for sparse samples of halos using $k\nn$ measurements when two-point measurements fail to detect a signal. Finally, we conclude, and discuss some aspects of the presentation in Sec. \ref{sec:conclusions}

\section{Formalism and measurement of cross-correlation in the nearest neighbor framework}
\label{sec:NNCDF}

\subsection{Formalism}
\label{sec:theory}
In this section, we lay out the formalism for describing spatial cross correlations between two datasets in terms of joint data counts in a given volume. We consider two sets of tracers - each set tracing an underlying continuous density field. Throughout this paper, we will assume that the tracers are distributed according to a local Poisson point process on the underlying fields, and that the Poisson parameter is proportional to the enclosed ``mass'' of the density field over the volume of interest. Therefore, fluctuations in the underlying fields are imprinted onto the fluctuations in the number counts of the tracers. The two sets of tracers will have non-zero cross-correlations when the over-densities and under-densities of the underlying fields coincide with each other on average -- the cross-correlation is the highest when the fluctuations in the two fields coincide exactly.

These cross-correlations will be imprinted on the \textit{joint} count distribution of the tracers of the two fields, \textit{i.e.} the probability of finding $k_1$ tracers from set $1$ \textit{and} $k_2$ tracers from set $2$, in a volume $V$ centered around random points in the volume of interest. As derived in Appendix \ref{sec:derivation}, the generating function, $P(z_1,z_|V)$, for the joint counts in volume $V$ is given by 
\eq{generating_function}{P\left(z_1,z_2|V\right) = \exp\Bigg[&\sum_{k_1=0}^\infty \sum_{k_2=0}^\infty \frac{\bar n_1^{k_1}(z_1-1)^{k_1}}{k_1!}\frac{\bar n_2^{k_2}(z_2-1)^{k_2}}{k_2!}\nonumber \\ &\times \int_V d^3\mbr_1...d^3\mbr_{k_1}d^2\mbr_1^\prime ...d^3\mbr_{k_2}^\prime \xi^{(k_1,k_2)}\Bigg]\, , }
where $\bar n_i$ represent the mean number density of each set of tracers, and $\xi^{(k_1,k_2)}$ represents the connected correlation function (in terms of the continuous fields) defined with $k_1$ factors of field $1$ and $k_2$ factors of field $2$. For example $\xi^{(2,0)}$ represents the two-point auto-correlation function of field $1$, $\xi^{(0,2)}$ represents the two-point auto-correlation function of field $2$, and $\xi^{(1,1)}$ represents the two-point cross correlation between the two fields. Note that if the two fields are uncorrelated, or statistically independent, \textit{i.e.}, $\xi^{(k_1,k_2)}$ is non-zero only when either $k_1=0$ or $k_2=0$, the generating function factorizes into two independent generating functions - one for each set of tracers: $P(z_1,z_2|V) = P_1(z_1|V)P_2(z_2|V)$. Also note that while the volume $V$ in Eq, \ref{eq:generating_function} can have any arbitrary shape in general, we will focus on the scenario where it corresponds to a sphere of radius $r$.

The probability $\mathp(k_1,k_2|V)$ of finding exactly $k_1$ counts of data from set $1$, and $k_2$ counts of data from set $2$ in volume $V$ are given by various derivatives of the generating function with respect to the dummy variables $z_1$ and $z_2$:
\eq{counts_from_generating_function}{\mathp(k_1,k_2|V) = \frac{1}{k_1!}\frac{1}{k_2!}\Bigg[\Bigg(\frac{\dd}{\dd z_1}\Bigg)^{k_1}\Bigg(\frac{\dd}{\dd z_2}\Bigg)^{k_2}P(z_1,z_2|V)\Bigg]_{z_1,z_2=0}\, .}
From the forms of Eq. \ref{eq:generating_function} and \ref{eq:counts_from_generating_function}, it is evident that, for all values of $k_1,k_2$, the probabilities $\mathp(k_1,k_2|V)$ are related to all possible connected $N$-point functions that can be defined from fields $1$ and $2$. This includes all possible cross terms -- $\xi^{(k_1,k_2)}$ for non-zero values of both $k_1$ and $k_2$. 

Using the same formalism, it is also possible to write down the generating function $C(z_1,z_2|V)$ for the joint \textit{cumulative} counts, \textit{i.e.} the probability of finding more than $k_1$ tracers from set $1$ \textit{and} more than $k_2$ tracers in volume $V$.
\eq{cumulative_generating_function}{C(z_1,z_2|V) = \frac{1-P_1(z_1|V)-P_2(z_2|V)+P(z_1,z_2|V)}{(1-z_1)(1-z_2)}\,,}
where $P_i(z_i|V)$ represent the generating function for the counts of each individual set of tracers. Note that in the absence of cross-correlations, \textit{i.e.} when $P(z_1,z_2|V) = P_1(z_1|V)P_2(z_2|V)$, this generating function also factorizes into a product of the generating functions for the cumulative counts for each distribution individually:
\eq{cumulative_disjoint}{C(z_1,z_2|V) &= \frac{1-P_1(z_1|V)-P_2(z_2|V)+P_1(z_1|V)P_2(z_2|V)}{(1-z_1)(1-z_2)}\nonumber \\ &= \Bigg(\frac{1-P_1(z_1|V)}{1-z_1}\Bigg)\Bigg(\frac{1-P_2(z_2|V)}{1-z_2}\Bigg) \nonumber \\ &= C_1(z_1|V)C_2(z_2|V)\, .}
Just as in Eq. \ref{eq:counts_from_generating_function}, one can compute the individual terms $\mathp(>\ns k_1, >\ns k_2|V)$ for any value of $k_1, k_2$ from the derivatives of $C(z_1,z_2|V)$:
\eq{cumulative_counts_from_gf}{\mathp(>\ns k_1,> \ns k_2|V) = \frac{1}{k_1!}\frac{1}{k_2!}\Bigg[\Bigg(\frac{\dd}{\dd z_1}\Bigg)^{k_1}\Bigg(\frac{\dd}{\dd z_2}\Bigg)^{k_2}C(z_1,z_2|V)\Bigg]_{z_1,z_2=0}\, .}
There are two issues to note here that will be relevant throughout the paper: first, the joint cumulative probabilities $\mathp(>\ns k_1,> \ns k_2|V)$ are sensitive to terms which capture the cross-correlation between fields $1$ and $2$, \textit{i.e.} $\xi^{(k_1,k_2)}$ for $k_1 \neq 0, k_2\neq 0$, as well as terms which capture the clustering of the two fields individually, \textit{i.e.} terms like $\xi^{(k_1,0)}$ and $\xi^{(0,k_2)}$. Second, in the case where the underlying fields are uncorrelated, 
\eq{uncorrelated_cumulative}{
\mathp(>\ns k_1,> \ns k_2|V) = \mathp_1(>\ns k_1|V)\mathp_2(>\ns k_2|V)\; ,}
and therefore, any deviation from this condition can be treated as a measure of the degree of correlation between the two fields under consideration. We will return to the implications of this factorization in Sec. \ref{sec:measurement}, but it is worth noting here that the term ``uncorrelated'' as used here is a more stringent criterion than just the absence of two-point cross-correlation, as often used in the literature. Since the $k\nn$ measurements are sensitive to not only the two-point cross-correlation, but all higher-order terms, ``uncorrelated'' in this context implies complete statistical independence of the two distributions. 

Having set up the formalism to show how these joint cumulative probabilities of the tracers capture the correlations between two different underlying fields, in the next section we explore how these probabilities can be measured efficiently from given datasets, through the \textit{joint} $k$ Nearest Neighbor Cumulative Distribution Functions ($k\nn$-$\cdf$).

\subsection{Measurement of joint $k\nn$-$\cdf$}
\label{sec:measurement}

To set up the measurements for the joint $k\nn$-$\cdf$, we populate the volume of interest, $V_{\rm tot}$, typically representing the simulation volume throughout this paper, with $N_R$ random points. Using higher numbers of random points to sample the volume leads to lower measurement noise in the $\cdf$. Let the two sets of tracers for which we want to compute the joint clustering have $N_{D_1}$ and $N_{D_2}$ points each, distributed over the volume under consideration. We build two separate $k$-d trees (see \textit{e.g.} \citet{Wald2006}) from each set of particle positions, and then use these trees to find and store the distance to the $k$ nearest neighbor data points from each random point. There are publicly available tree codes (\textit{e.g.} \textsc{Scipy}'s \textsc{cKDTree} implementation, and \textsc{Julia}'s \textsc{NearestNeighbor.jl}\footnote{https://github.com/KristofferC/NearestNeighbors.jl} library) that can be used to efficiently carry out this calculation in $N \log N$ operations.

First, consider the case of $k=1$ for both datasets - the distance from the randoms to the \textit {first} nearest neighbor data point in each set. For each random point, therefore, we can associate two distances - one to the nearest neighbor data point from the first set, and the other to the nearest neighbor data point from the second set. Now, for every random point, we choose the larger of the two distances. These distances are then sorted to get the empirical Cumulative Distribution Function (CDF) of the distances chosen in this manner. We will refer to this distribution as the \textit{joint} Nearest Neighbor CDF, $\cdf_{1,1}$. This joint $\nn$-$\cdf$ can be interpreted in the following way: at a fixed radius $r$ (or the corresponding spherical volume $V=4/3\pi r^3$), the value of the CDF represents the fraction of spheres for which the distance to the nearest neighbor data point in \textit{both} sets are smaller than $r$. It is, therefore, equivalent to the fraction of spheres of radius $r$ which contains at least one data point of the first dataset \textit{and} the second dataset, \textit{i.e.},
\eq{connection_00}{\cdf_{1,1}(r) = P(>\ns 0, >\ns 0|V) \, .}
It is easy to generalize this argument to show that 
\eq{connection}{\cdf_{k_1,k_2}(r) = P\big(>\ns k_1-1, >\ns k_2-1|V\big)\, ,}
where the $\cdf_{k_1,k_2}$ is the CDF computed as outlined above, but by considering the distance to the $k_1-$th nearest neighbor data point from set $1$ and the distance $k_2-$th nearest neighbor data point from set $2$ from every random point. Through the formalism in Sec. \ref{sec:theory}, therefore, the \textit{joint} $k\nn$-$\cdf$ as defined here are sensitive to all the connected $N$-point correlation functions that can be formed from the two underlying fields from which the data points are sampled.

It is also possible, through the measurements of the nearest neighbor distributions, to isolate those parts of $P(>\ns k_1, >\ns k_2|V)$ which depend \textit{only} on the cross-correlation of the two fields, and not on the clustering of the two fields individually. To do this, we use Eq. \ref{eq:uncorrelated_cumulative} to obtain the prediction for the measurement of $P(>\ns k_1, >\ns k_2|V)$ in the uncorrelated scenario - in this case, it is just a product of the individual probabilities, $\mathp_1(>\ns k_1|V)$ and $\mathp_2(>\ns k_2|V)$. As shown in \citet{Banerjee_Abel}, $\mathp (>\ns k|V) = \cdf_{k+1}(r)$, where $\cdf_{k_i}(r)$ represents the Empirical CDFs for the distances to the $k$-th nearest neighbor data points computed separately for each dataset (labeled by $i$). Since these individual distances from the random points in the box to the $k$-th nearest neighbor of each dataset are measured anyway as one of the steps toward building up the joint $\cdf$, there is no significant additional resources needed to sort the distances and compute the individual $\cdf$s. Therefore, by subtracting the product of the relevant $k_i$-th nearest neighbor distributions of each set of tracers from the \textit{joint} nearest neighbor measurements, we are left with the piece $\psi^{(k_1,k_2)}(r)$ that parameterizes the cross correlation of the two fields:
\eq{xi_tilde}{\psi^{(k_1,k_2)}(r) = \cdf_{k_1,k_2}(r) - \cdf^{(1)}_{k_1}(r)\cdf^{(2)}_{k_2}(r)\, ,}
where the superscripts on the RHS indicate the set of tracers for which the CDF has been computed. Higher absolute values of $\psi^{(k_1,k_2)}(r)$, for fixed amplitudes of clustering in the individual fields, indicate higher levels of spatial correlation between fields $1$ and $2$ at scale $r$ - positive values indicate positive correlations, while negative values indicate anti-correlations in the two fields. If $\psi^{(k_1,k_2)}(r)=0$, the two sets of tracers are uncorrelated. We reiterate that this implies not only an absence of the two-point cross-correlations, but a complete statistical independence of the two fields, as captured by all possible combinations of $N$-point functions of the two fields.

\begin{figure*}
	\includegraphics[width=0.9\textwidth]{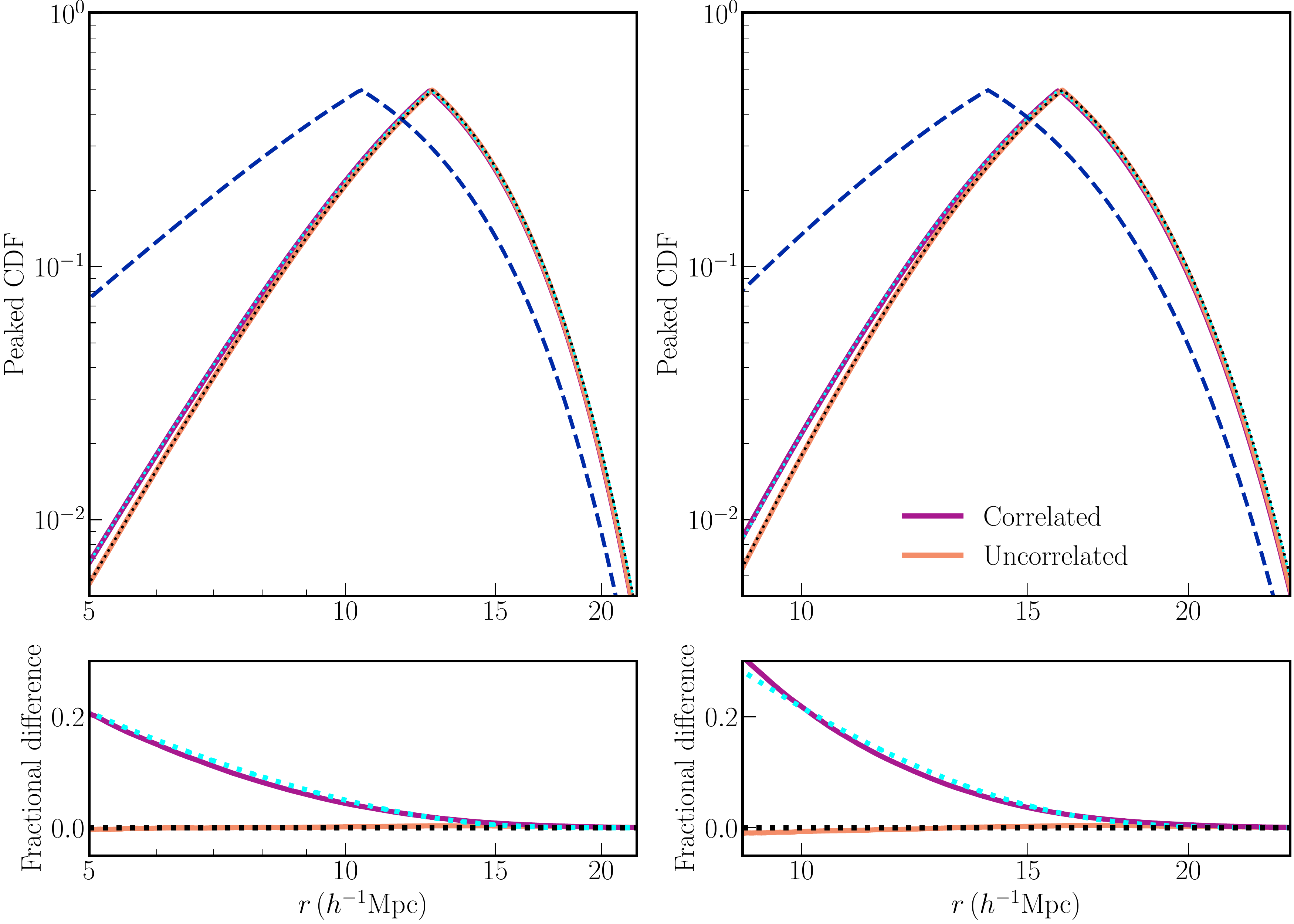}
	\caption{\textit{Top panels}: Solid lines represent the peaked CDF of the \textit{joint} $1\nn$ (top-left panel) and \textit{joint} $2\nn$ (top-right panel) distributions for two correlated (darker lines) and two uncorrelated (lighter lines) sets, each composed $2\times 10^5$ tracers of a Gaussian random field over a $(1\hgpc)^3$ volume (see Sec. \ref{sec:Gaussian} for details). Dotted lines indicate the theoretical expectations for these measurements. The dashed lines represent the $1\nn$(left panel) and $2\nn$ (right panel) measurements for only one of the tracer sets, shown as a reference. \textit{Bottom panels}: We plot the fractional differences of the predictions and measurements from the upper panels with respect to the analytic predictions for the uncorrelated sets for the joint $1\nn$ (bottom-left panel) and joint $2\nn$ (bottom-right panel) distributions. The differences in the joint CDFs between the correlated and uncorrelated datasets are especially clear on small scales, and match well with the analytic expectations. The different scales plotted on the left and right panels indicate the range of scales over which the distributions are well measured with the choice of measurement parameters mentioned in Sec. \ref{sec:Gaussian}.}
	\label{fig:gaussian}
\end{figure*}

It is worth noting here the computational expense associated with these measurements. We focus on a typical example where we use $2\times 10^5$ tracers in each set, distributed over a $(1\hgpc)^3$ volume. The number of randoms, from the nearest neighbor distances are measured, was $4\times10^6$ distributed uniformly over the same volume. We measure $\psi^{(k_1,k_2)}(r)$ for $k_1=k_2=k\in \{1,2,3,4\}$. The entire measurement takes $\sim 45$ seconds on a single core. The tree construction for each sample only takes less than a second, while each tree query takes $\sim 18$ seconds. Since the largest cost for typical parameter choices is associated with the tree search for the nearest neighbors from the randoms, parallelizing this part can further reduce runtime --- \textsc{Scipy}'s \textsc{cKDTree} implementation, for example, already allows for this through the \texttt{njobs} flag. We also note that the computational expense does not scale strongly with the value of $k$, especially for the range of values of $k$ we use in this study - the runtime with $k_1=k_2=1$ is roughly similar to that in the example above. 

In the rest of the paper, we will generally consider datasets with roughly equal number densities. Even if the datasets have different number densities originally, such as dark matter particles and halos in a simulation, we will usually downsample the denser one to match the number density of the sparser sample. Having matched the number densities, we will generally consider $\cdf_{k,k}(r)$ in our analysis, where $k_1=k_2=k$. However, it should be noted that these are merely choices driven by considerations of the scales of interest, as well as computational time. One could also consider the joint $\nn$ distributions when $k_1\neq k_2$. This may be especially relevant and appropriate when the two datasets have very different number densities.

\section{Correlated Gaussian fields}
\label{sec:Gaussian}

In this section, we demonstrate the application of the measurement method outlined in Sec. \ref{sec:NNCDF} to the tracers of two correlated Gaussian fields. For completely Gaussian fields, the expression for the generating function in Eq. \ref{eq:generating_function} can be truncated by only retaining terms up to the 2-point correlation functions --- all higher order terms can be set to $0$. Writing this out explicitly, we have
\eq{Gaussian_generating_function}{P\left(z_1, z_2|V\right) = \exp\Bigg[& \bar n_1(z_1-1)V+\bar n_2(z_2 - 1) V \nonumber \\ &+ \frac 1 2 \bar n_1^2(z_1 - 1)^2\bar \xi^{(2,0)}_V + \frac 1 2 \bar n_2^2(z_2 - 1)^2\bar \xi^{(0,2)}_V\nonumber \\ &+\bar n_1\bar n_2 (z_1-1)(z_2-1)\bar \xi^{(1,1)}_V\Bigg]\, ,}
where 
\eq{xi_bar_definition}{\bar \xi^{(k_1,k_2)}_V = \int_V d^3\mbr_1...d^3\mbr_{k_1}d^2\mbr_1^\prime ...d^3\mbr_{k_2}^\prime \xi^{(k_1,k_2)}\, ,}
and $\bar n_i$ represent the mean number density of each set of tracers.  In this notation $\bar \xi^{(2,0)}_V$ and $\bar \xi^{(0,2)}_V$ represent the two point auto-correlation functions of fields $1$ and $2$ respectively, integrated over volume $V$. $\bar \xi^{(1,1)}_V$ represent the 2-point cross correlation of fields $1$ and $2$ integrated over volume $V$. Note that the integrated cross correlation can be negative if the two fields are anti-correlated. Using the fact that for a single Gaussian field (see \citet{Banerjee_Abel} for details)
\eq{gaussian_single}{P(z|V) = \exp \Bigg[\bar n (z -1)V + \frac 1 2 \bar n^2 (z-1)^2\bar \xi^{(2)}_{V}\Bigg] \, ,}
the full expression for generating function of the joint cumulative counts in Eq. \ref{eq:cumulative_generating_function}, $C(z_1,z_2|V)$, can be written down, and the individual $\mathp (>\ns k_1,>\ns k_2|V)$ can be evaluated from the derivatives. The functional forms of the first few terms, are shown in Appendix \ref{sec:Gaussian_expressions}, and we will use these expressions to compare with the measurements outlined below.

For the measurements, we consider two $(1\hgpc)^3$ simulations with $512^3$ CDM particles run from $z=99$ to $z=4$ at the \textit{Planck} best-fit cosmology \citep{2020A&A...641A...6P}. The two simulations have different realizations of the initial power spectrum, and therefore the final density fields should not be spatially correlated. At the redshift under consideration, the matter field is still sufficiently close to Gaussian for the purposes of this exercise. We randomly downsample the set of simulation particles from the first realization down to $2\times 10^5$ tracer particles. We then choose a different set of $2 \times 10^5$ particles from the same realization, ensuring that the same particle does not end up in both datasets. We first measure the $k\nn$-$\cdf$ for only the first set of $2\times 10^5$ particles for $k=1$ and $k=2$. These measurements are represented by the dashed lines on the upper left and right panels of Fig. \ref{fig:gaussian}, respectively. We then perform the \textit{joint} nearest neighbor CDF measurements for $k_1=k_2=1$ and $k_1=k_2=2$ using both sets of particles. These measurements are represented by darker solid lines in the upper left and right panels of Fig. \ref{fig:gaussian}. Since both sets have been sampled from the same realization, we expect them to be spatially correlated. Next, we choose a random set of $2\times 10^5$ particles from the second realization, and measure the joint nearest neighbor measurements for $k_1=k_2=1$ and $k_1=k_2=2$ between this set of particles and one of the set of downsampled particles from the first realization. These measurements are represented by the lighter solid line in the top panels (left and right) of Fig. \ref{fig:gaussian}. Since the two sets of particles in this latter case are from two different realizations, they are expected to be spatially uncorrelated.

\begin{figure*}
	\includegraphics[width=0.9\textwidth]{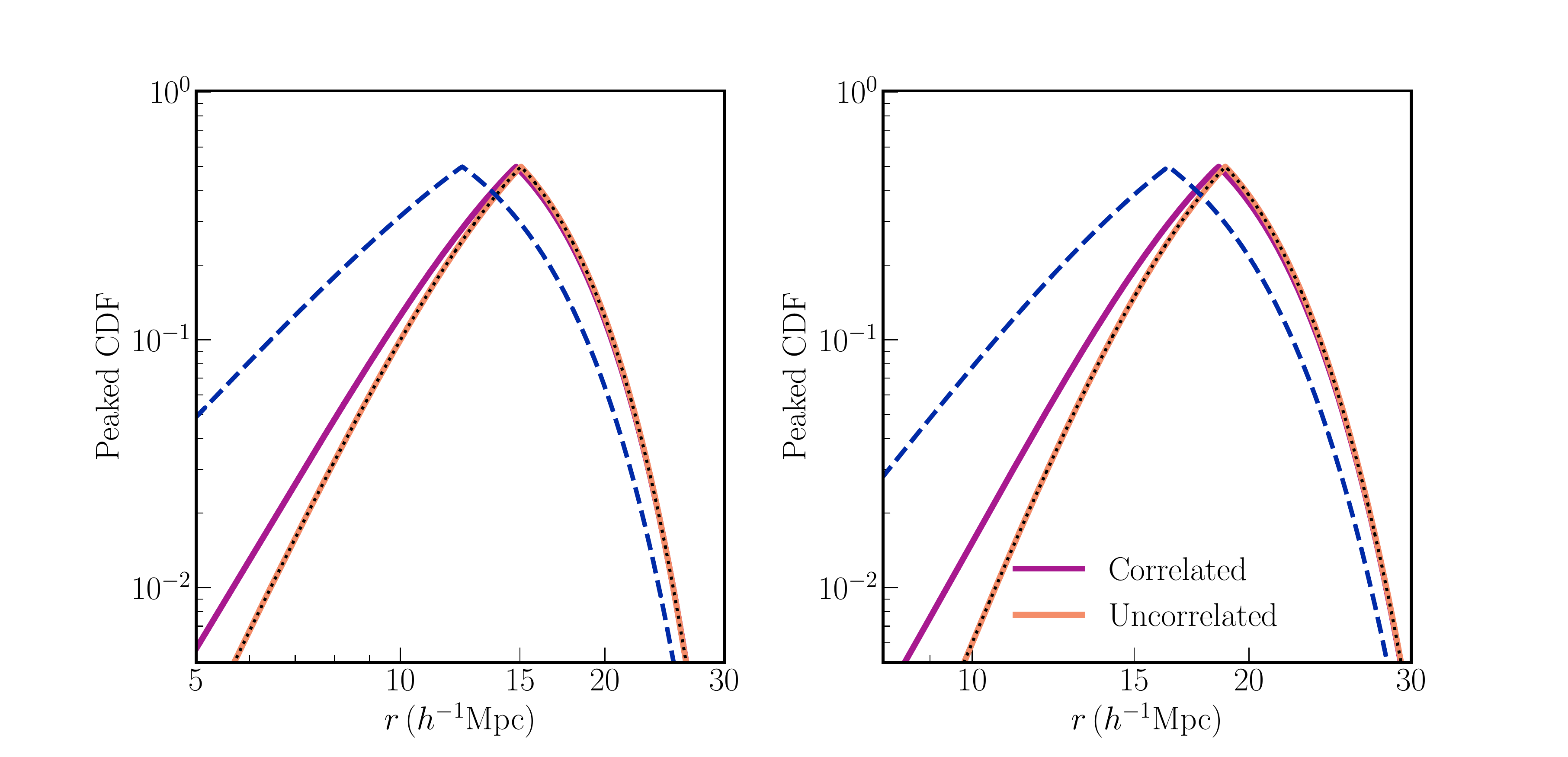}
	\caption{Solid lines represent the peaked CDF of the \textit{joint} $1\nn$ (left panel) and \textit{joint} $2\nn$ (right panel) distributions for two correlated (darker lines) and two uncorrelated (lighter lines) sets of simulation particles at $z=0$, when the matter field is highly nonlinear on small scales. Each set has $10^5$ particles downsampled from a $(1\hgpc)^3$ simulation with $512^3$ particles. The dashed line on each panel represents the first and second nearest neighbor peaked CDF for a single set of particles for reference. The dotted line in each panel represents the expectation for the joint $1\nn$ and $2\nn$ CDFs of two uncorrelated sets of particles given the measurements of their individual $1\nn$ and $2\nn$ distributions. Deviations from this dotted line in each panel represents the degree of cross-correlation between the datasets. The range of scales on each panel represents the range over which the distributions are well measured for the specific choice of parameters. See text for more details.}
	\label{fig:lss}
\end{figure*}

Since the fields at $z=4$ are still close to Gaussian, we use the \textsc{Colossus} software\footnote{http://www.benediktdiemer.com/code/colossus/} to compute the variance of the matter field fluctuations as a function of scale $\sigma(r)$, as expected from linear perturbation theory. We note that \textsc{Colossus} by default uses the Eisenstein-Hu approximation \citep{1998ApJ...496..605E} for the matter power spectrum, and is accurate at the level of $\lesssim 5\%$. However, this is sufficient for the purposes of this exercise: to demonstrate differences in the joint nearest neighbor distributions of correlated and uncorrelated samples. The output from \textsc{Colossus} is used to evaluate the analytic expectations for the joint $\nn$-$\cdf$s for both the correlated and uncorrelated sets, using Eqs. \ref{eq:gaussian11} and \ref{eq:gaussian22a}. These predictions are plotted using the dotted lines in the two upper panels of Fig. \ref{fig:gaussian}. In the bottom panel of Fig. \ref{fig:gaussian}, we plot the fractional difference in the various predicted and measured $\cdf$, using the analytic prediction for the uncorrelated case as the baseline. The bottom-left panel is for $1\nn$ measurements, and the bottom-right panel is for the $2\nn$ measurements. The color scheme is the same as in the top panel. It is clear, especially at the smaller scales that there is a difference in the nearest neighbor distributions between the two scenarios, and that these differences are captured correctly by the analytic predictions. Note that in the uncorrelated case, the prediction is that the joint $\nn$-$\cdf$s are simply a product of the individual 
$\nn$-$\cdf$s of each dataset. 

The cosmological matter density field, and therefore the distribution of tracer particles, is even closer to a true Gaussian random field at higher redshifts. However, since the amplitude of the fluctuations are also lower at these higher redshifts, it is difficult to visualize the differences between the correlated and uncorrelated scenarios in Fig. \ref{fig:gaussian}. As we will see in the next section, the differences between correlated and uncorrelated datasets become clearer when the joint nearest neighbor measurements are applied to fully nonlinear fields, like the matter density field at $z=0$.

\section{Correlations in nonlinear fields}
\label{sec:LSS}

We now focus on measuring joint $k\nn$-$\cdf$ and cross-correlations, $\psi^{(k_1,k_2)}(r)$, in cosmological fields with nonlinear clustering, \textit{i.e} clustering of matter and dark matter halo at low redshifts. Unlike in the Gaussian example presented in Sec. \ref{sec:Gaussian}, the number of terms which are important in the generating functions, both for the individual counts, and for the joint counts, are not known beforehand. This implies that it is not always possible to write down an analytic expression for the counts in terms of the $N$-point correlation functions. However, the measurements of the individual and joint cumulative counts using the nearest neighbor measurements can be be performed exactly as outlined in Sec. \ref{sec:measurement}.

To demonstrate the measurement of cross-correlations in these nonlinear fields, we follow the general outline of Sec. \ref{sec:Gaussian} and consider the following example: we take a $(1\hgpc)^3$ simulation at $z=0$ with $512^3$ particles and choose a random subset of $10^5$ particles. We measure the nearest neighbor distributions for this single set of particles and plot the peaked $\cdf$ using the blue dashed lines in the Fig. \ref{fig:lss} - the left panel shows the $k_1=k_2=1$ nearest neighbor distribution while the right panel shows the $k_1=k_2=2$ nearest neighbor distribution. Next, we choose another random subset of $10^5$ particles from the same simulation - we do not use any of the particles that were part of the first sample. Since both sets of particles are drawn from the same simulation, and therefore trace the same underlying density field, they should be fully correlated, modulo sampling noise. We now measure the \textit{joint} nearest neighbor distribution for these two sets of tracers. The results of the $k_1=k_2=1$ and $k_1=k_2=2$ joint nearest neighbor peaked $\cdf$ are represented by the darker solid lines in the left and right panels of Fig. \ref{fig:lss}, respectively.

We then take another simulation with the same resolution and at the same cosmology, but with a different random realization of the initial density field. We randomly select $10^5$ particles from this simulation. Since the initial modes for this realization and that of the first simulation are different, the final density fields do not align with each other, and therefore, the tracers are also not expected to have any statistical correlations in their clustering. We measure the joint nearest neighbor distributions for the two sets of $10^5$ particles from the two realizations. The results of the $k_1=k_2=1$ and $k_1=k_2=2$ joint nearest neighbor peaked $\cdf$ are represented by the lighter solid lines in the left and right panels of Fig. \ref{fig:lss}, respectively.

Unlike in the Gaussian case, there is no simple analytic expectation for the joint $k\nn$-$\cdf$ in the most general correlated case. However, it is still possible to make a prediction for the joint $k\nn$-$\cdf$ for uncorrelated samples based on the measurements of the \textit{individual} $k\nn$-$\cdf$ using Eq. \ref{eq:uncorrelated_cumulative}. This expectation is plotted using the dotted black lines in both panels of Fig. \ref{fig:lss}, and agrees with the direct measurements for the uncorrelated samples on all scales displayed in the plot. The difference between the correlated and uncorrelated samples are especially pronounced on smaller scales where the clustering is stronger, but the difference persists out to the largest scales that we measure, and will become evident when considering $\psi^{(k_1,k_2)}(r)$ in the following subsections.

Having demonstrated the measurement of joint $k\nn$-$\cdf$ measurements for nonlinear cosmological fields, we explore two applications of these measurements below - one in the high signal-to-noise regime, where we can use cross-correlation measurements to infer cosmology, and one in the low signal-to-noise regime, where we focus simply on the detection of a signal. Given the scope of this paper, we will focus on the parts of the measurements, $\psi^{(k_1,k_2)}$, that capture cross-correlations in both applications.

\subsection{Parameter constraints using cross-correlations}
\label{sec:cosmo_constraints}

\begin{figure}
	\includegraphics[width=0.45\textwidth]{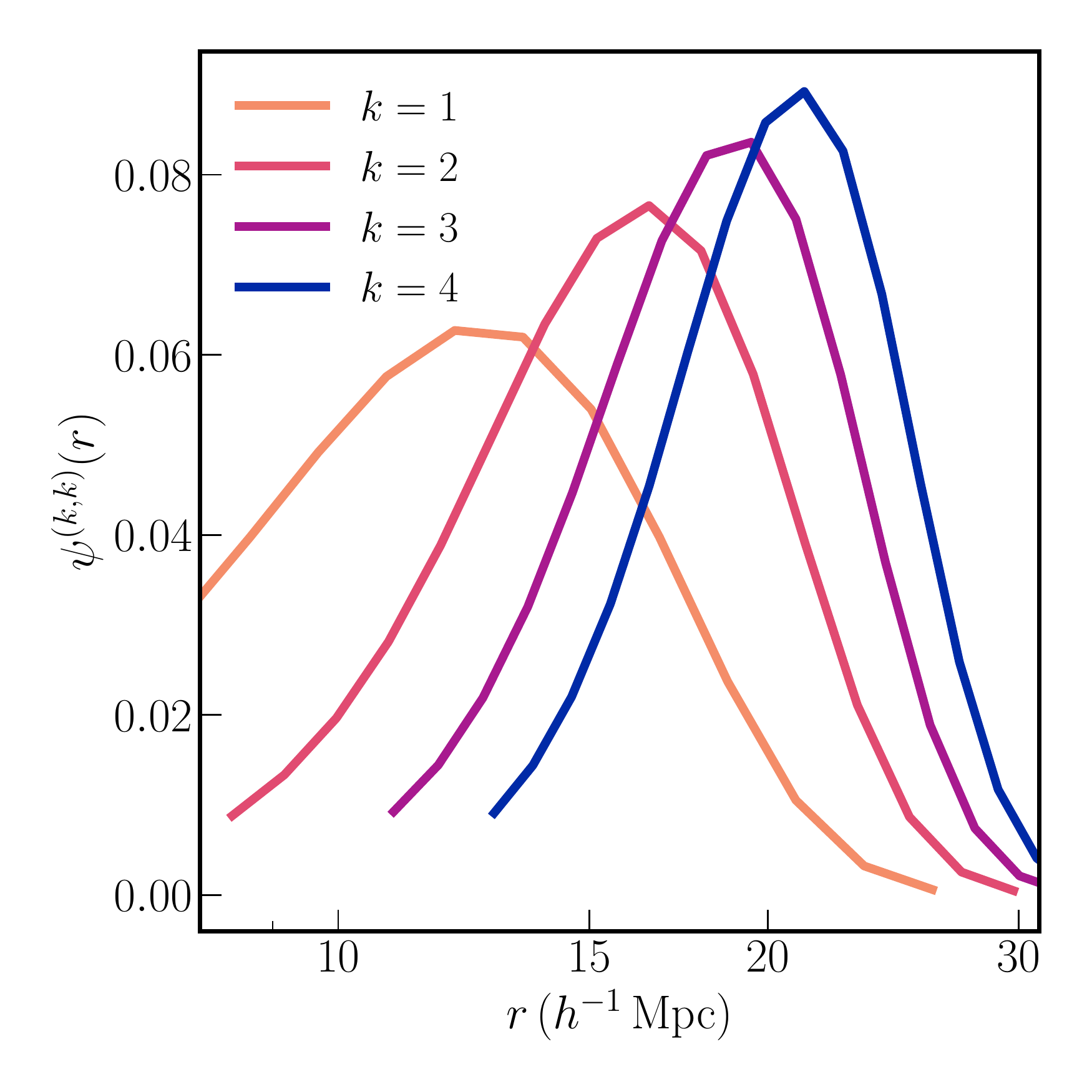}
	\caption{$\psi^{(k,k)}(r)$ (see Eq. \ref{eq:xi_tilde}), which measures the spatial correlation between two samples, for various $k$ measured from the $10^5$ most massive halos and $10^5$ randomly chosen particles from a $(1\hgpc)^3$ simulation at $z=0$. These measurements are used in the Fisher matrix calculations in Sec. \ref{sec:cosmo_constraints}. For each $k$ we plot the measurements over the range of scales that are used in the analysis for that particular $k$.}
	\label{fig:xi_prime}
\end{figure}

In this section, we use the Fisher matrix formalism to compare the information content, and sensitivity to the underlying cosmological parameters, of the two-point cross-correlation with that of the nearest neighbor method of computing cross-correlations. For this exercise, we use data from the \textsc{Quijote} suite of simulations\footnote{https://github.com/franciscovillaescusa/Quijote-simulations} \citep{2020ApJS..250....2V}. These simulations, run over different cosmologies, have a volume of $(1\hgpc)^3$ and use $512^3$ CDM particles for cosmologies without massive neutrinos, and $512^3$ CDM and $512^3$ neutrino particles for cosmologies with massive neutrinos.  We consider the cross-correlations of the $10^5$ most massive halos in the simulations with the underlying matter field at $z=0$. \citet{Banerjee_Abel} has demonstrated that nearest neighbor measurements of the auto clustering of these datasets are also more sensitive to cosmological parameters than two-point measurements --- here we demonstrate the same for only the cross-correlation piece. For the two-point cross-correlation, $\xi^{hm}$, we use \textsc{Corrfunc}\footnote{https://github.com/manodeep/Corrfunc} \citep{2020MNRAS.491.3022S, 10.1007/978-981-13-7729-7_1} to compute the results over $30$ bins between $8 \hmpc$ to $30 \hmpc$. For the nearest neighbor calculation, we downsample the simulation particles to $10^5$ randomly selected particles, and compute the joint nearest neighbor distributions for $k_1=k_2 =k \in \{1,2,3,4\}$ for these two datasets (halos and downsampled particles). We use $8\times 10^6$ random points over the simulation volume for the joint nearest neighbor calculations, using $16$ bins for each joint $k\nn$-$\cdf$, between $8 \hmpc$ to $30 \hmpc$ --- the same range used for the two-point cross-correlations. We use measurements of 
$\psi^{(k,k)}(r)$ in our analysis, which isolates the cross-correlations, rather than the direct measurements of the joint $k\nn$-$\cdf$ which are also sensitive to the clustering of each dataset individually. We further restrict the range of scales for each $k$ to ensure that the distributions are well measured. This is done by considering, for each $k$, only those range of scales for which the measurement of $\psi^{(k,k)}(r)$ is $>0.005$ at the fiducial cosmology of the \textsc{Quijote} suite. This choice ensures that the measurements are not affected by the noise associated with the number of randoms used in the calculation, nor by the statistical fluctuations in the tails of the distribution. We plot $\psi^{(k,k)}(r)$ for different $k$, as measured from one of the simulations in the \textsc{Quijote} suite in Fig. \ref{fig:xi_prime}. The scales for which measurements from an individual value of $k$ are used in the analysis can be clearly seen from the different ranges over which they are plotted in the figure.

\begin{figure}
	\includegraphics[width=0.45\textwidth]{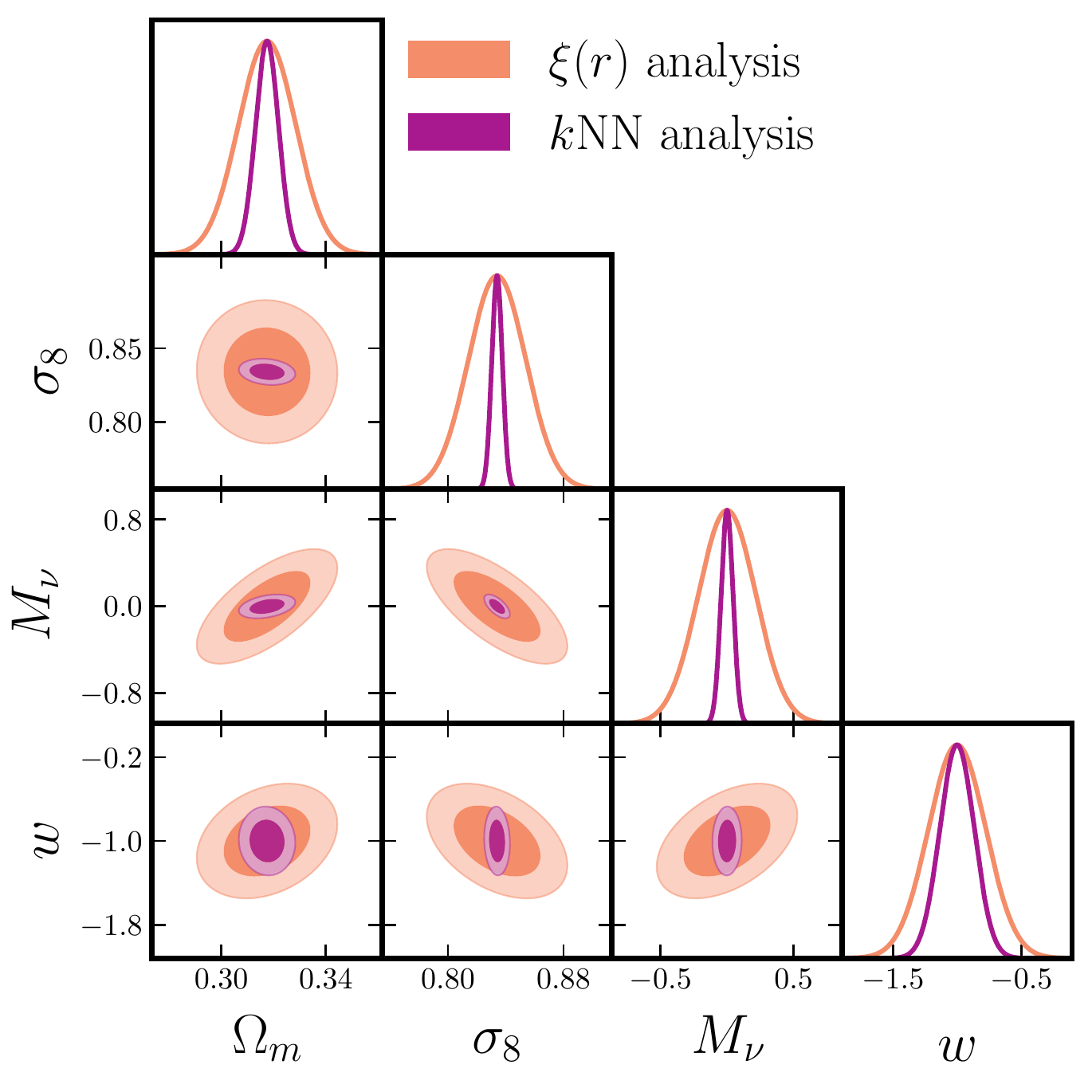}
	\caption{Constraints on various cosmological parameters from the Fisher analysis in Sec. \ref{sec:cosmo_constraints}. These are obtained from the cross-correlations of the $10^5$ most massive halos in the simulation volume ($(1\hgpc)^3$) with the matter distribution. There is a marked improvement in the constraints when the cross-correlations are measured through the nearest neighbor distributions ($k\nn$), compared to when measured through the two-point cross-correlation $\xi(r)$, over the same range of scales. The improvement is especially pronounced in some of the parameters, such as $\sigma_8$, and $M_\nu$. The individual constraints are listed in Table \ref{tab:part_constraints}.}
	\label{fig:cross_corr}
\end{figure} 

\begin{table}
	\centering
	\caption  {$1$-$\sigma$ constraints on cosmological parameters, from the $k\nn$ and $\xi(r)$ analysis of the cross-correlation of the matter field with the $10^5$ most massive halos in the box. The same range of scales are used in both analyses. }
	\label{tab:part_constraints}
	\begin{tabular}{c|c|c}
		\hline
		\hline
		${\rm Parameter}$ & $\sigma_{k\nn}$ & $\sigma_{\xi(r)}$\\
		\hline
		\hline
		$\Omega_m$ & 0.0061 & 0.0109 \\
		$\sigma_8$ & 0.0036 & 0.0191 \\
		$\Delta M_\nu$ & 0.0451 & 0.2161 \\
		$w$ & 0.1344 & 0.2239 \\
		\hline		
	\end{tabular}
\end{table}

We now briefly summarize the Fisher matrix formalism. We denote the data vector, either as measured through $\xi^{hm}(r)$ or through $\psi^{(k,k)}(r)$ as $\boldsymbol D$. The elements of the Fisher matrix are defined as
\eq{Fisher_definition}{\boldsymbol F_{\alpha\beta} = \sum_{i,j}\frac{\partial D_i}{\partial p_\alpha}\Big[\mathbf C^{-1}\Big]_{ij}\frac{\partial D_j}{\partial p_\beta}\,,}
where $\boldsymbol p$ represents the vector of cosmological parameters, and $\mathbf C$ represents the covariance matrix for the data vector. The Fisher matrix can be inverted to estimate how well various cosmological parameters are constrained by the particular data vector:
\eq{1sigma}{\sigma_\alpha = \sqrt{\big(\mathbf F^{-1}\big)_{\alpha\alpha}}\, .}
In this paper, the cosmological parameters considered are $\{\Omega_m, \sigma_8, M_\nu, w\}$. The covariance matrix is computed from the data vector computed over $1000$ realizations at the fiducial cosmology of the \textsc{Quijote} suite. The raw covariance matrix is computed as 
\eq{covmat_entries}{\mathbf C^\prime_{ij} = \Bigg\langle \bigg(D_i - \langle D_i\rangle\bigg) \bigg(D_i - \langle D_i\rangle\bigg)\Bigg \rangle \, ,}
where $\langle \boldsymbol D \rangle$ is the mean data vector averaged over the $1000$ realizations. We test for the stability of the Fisher analysis by checking that the condition number of the matrix is reasonable before inversion, and by checking that the distribution of values in each bins is roughly Gaussian around the mean. For each data vector considered, we use the Hartlap factor to correct for the fact that a finite number of realizations are used to estimate the covariance matrix \citep{2007A&A...464..399H}:
\eq{Hartlap}{\mathbf C^{-1} =  \frac{n-p-2}{n-1}\left( \mathbf C^\prime\right)^{-1}\, ,}
where $n$ represents the number of realizations, while $p$ represents the length of the data vector. This corrected covariance matrix is used in Eq. \ref{eq:Fisher_definition}. The \textsc{Quijote} suite is designed to allow for the derivatives of the data vectors with respect to the cosmological parameters to be easily evaluated, since the cosmologies are changed by one parameter at a time. We average over $100$ realizations at each of the derivative cosmologies to compute the data vector derivatives required in Eq. \ref{eq:Fisher_definition}.

\begin{figure*}
	\includegraphics[width=0.9\textwidth]{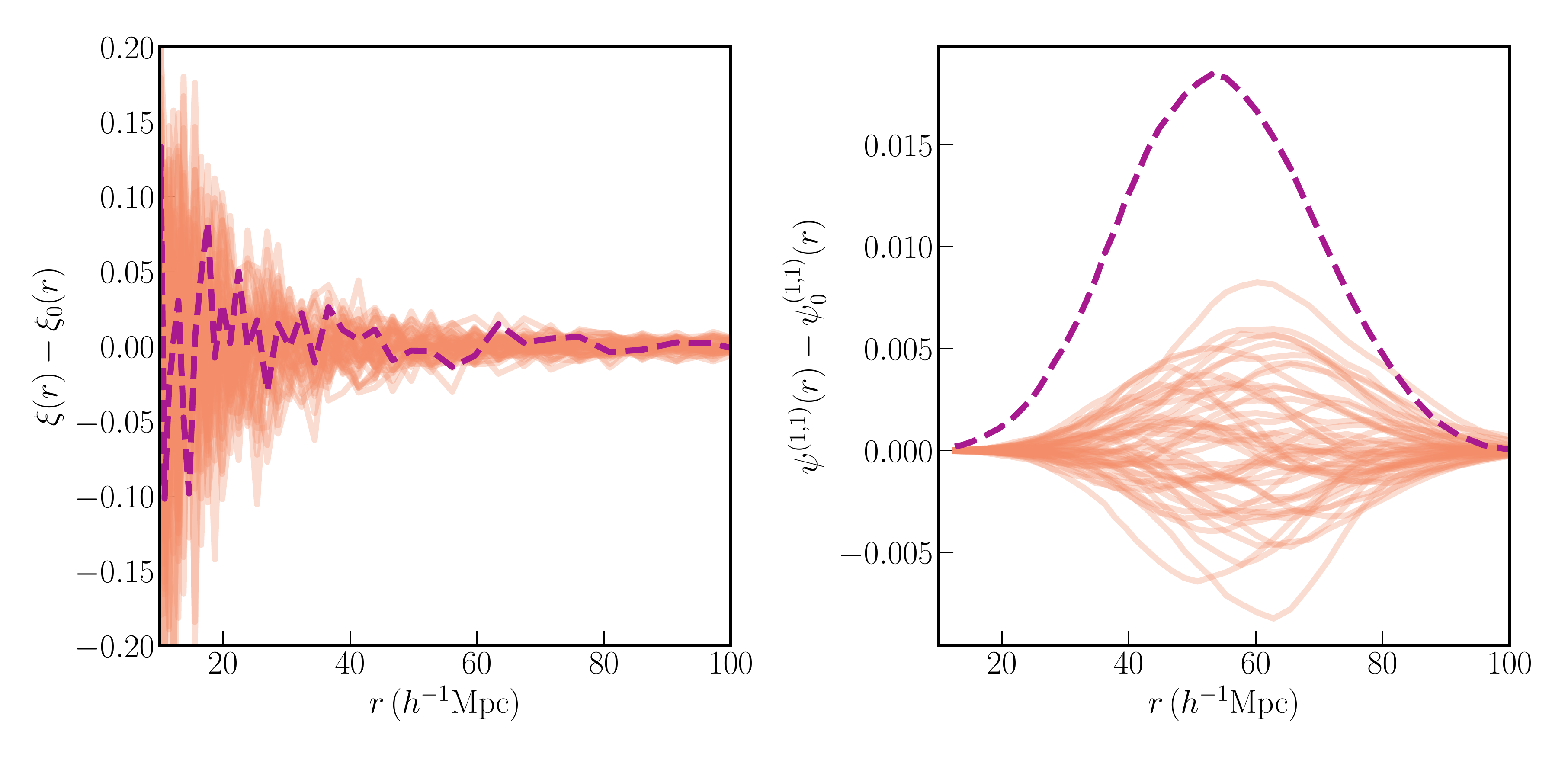}
	\caption{\textit{Left panel:} Difference of the two-point cross correlation measurements of two sets of dark matter halos ($1000$ halos each) from the mean of $1000$ such measurements where the two sets are spatially uncorrelated (drawn from different realizations). The lighter solid lines represent the difference for $50$ of these uncorrelated samples, meant to serve as a visual measure of the spread in the measurements when there are no true correlations. The darker dashed line represents the measurement of the same quantity in the case when the two sets of halos are from the simulation, and therefore, correlated. \textit{Right panel}: Same measurements as in the left panel, but using $\psi^{(1,1)}(r)$ (see Eq. \ref{eq:xi_tilde}) to measure cross-correlations instead of the two-point cross-correlation. Using $\psi^{(1,1)}(r)$, the correlated measurement is clearly separated from the uncorrelated ones. See Sec. \ref{sec:sparse samples} for more details.}
	\label{fig:ratio}
\end{figure*}

\begin{figure}[ht]
	\includegraphics[width=0.45\textwidth]{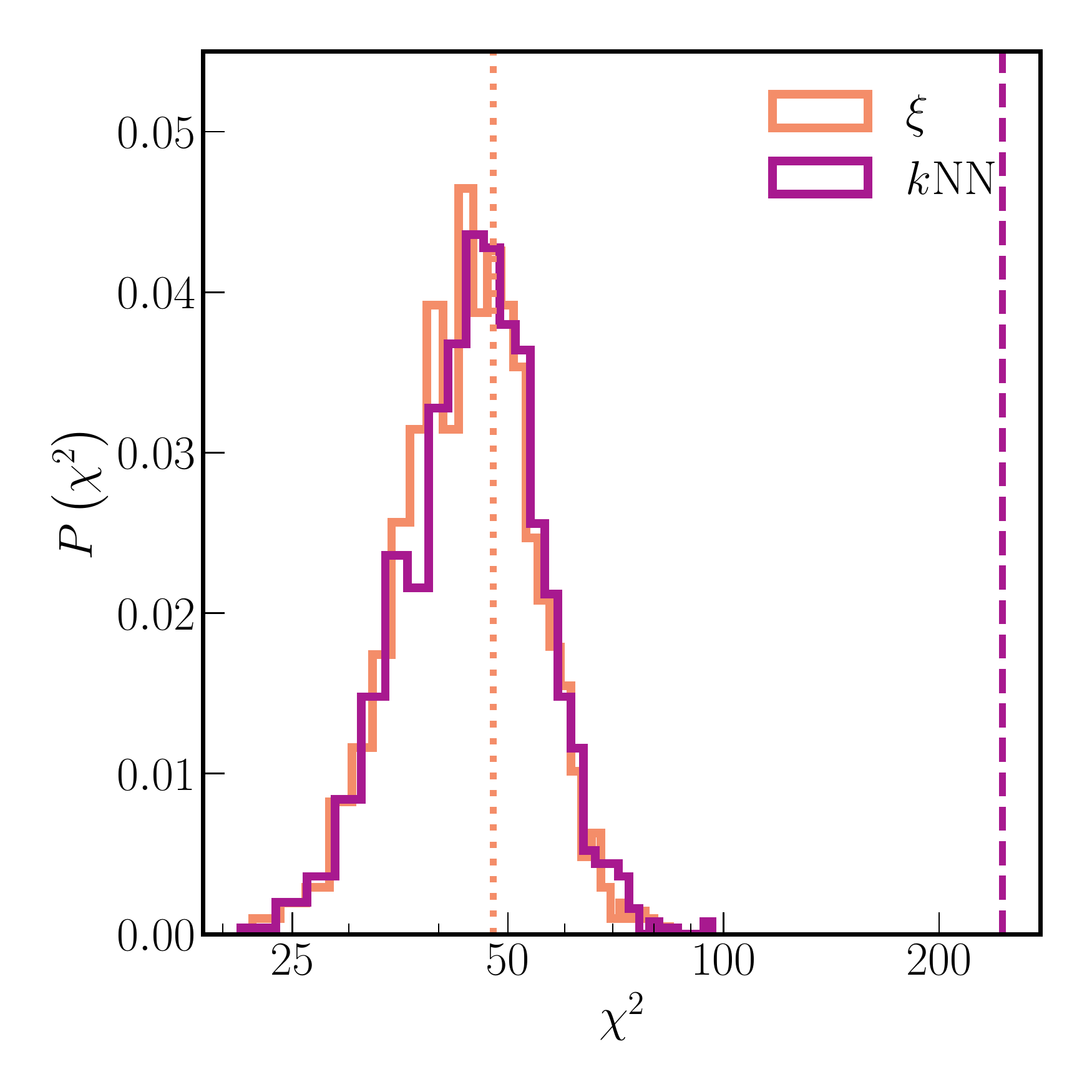}
	\caption{The solid lines represent the binned distribution of $\chi^2$ values for $1000$ measurements of cross-correlations between two samples of halos ($1000$ halos each from a $(1\hgpc)^3$ volume) which are spatially uncorrelated. The lighter shaded line represents the distribution when cross-correlations are measured through the two-point function ($\xi$), while the darker line represents the distribution when $k\nn$ measurements ($\psi^{(1,1)}$) are used to measure the cross-correlation. The dotted line represents the value of $\chi^2$ in the case when the two halo samples are spatially correlated, and when the cross correlation is measured through $\xi$. The dashed line represents the $\chi^2$ value when the cross-correlation of these samples is measured via the nearest neighbor distribution. The cross-correlation is clearly detected in the latter measurement, as seen by the $\chi^2$ value being far to the right of the distribution for the uncorrelated samples ($p-$value $<10^{-3}$). The two-point measurement, on the other hand, fails to detect a statistically significant correlation.}
	\label{fig:chi2}
\end{figure}

The results of the Fisher analysis, in terms of constraints on various cosmological parameters, and their covariance are presented in Fig. \ref{fig:cross_corr}. The constraints on individual parameters are also summarized in Table \ref{tab:part_constraints}. The lighter contours and posteriors in Fig. \ref{fig:cross_corr} represent the results from the analysis of the two-point cross-correlation, $\xi(r)$ while the darker contours and posteriors represent the results from the nearest neighbor ($k\nn$) analysis. We find that the nearest neighbor cross-correlations are more sensitive to cosmological parameters compared to the two-point cross-correlations. This is true for all the cosmological parameters considered here, but the improvement in constraints is especially pronounced for $\sigma_8$ and $M_\nu$. The degeneracy directions are also somewhat different in the two cases, which suggests further improvements in parameter constraints if the two measurements are combined. The overall improvements seen in Fig. \ref{fig:cross_corr} and Table \ref{tab:part_constraints} are not surprising, since we have shown that the two-point cross-correlations are a subset of all the terms that contribute to the measurements of $\psi^{(k,k)}$. Of course, a full analysis would consider both the cross-correlations of the matter field and halos, as well as the auto-clustering of each sample. Since the nearest neighbor measurements are considerably more sensitive to each of these (see \cite{Banerjee_Abel} for discussions on the auto-clustering), we can conclude that nearest neighbor measurements is a promising tool for all clustering measurements in cosmological analyses.

We now discuss a few caveats about the analysis presented above. First, it worth noting that the purpose of this exercise is to show the relevant improvement in cosmological parameter constraints when using $k\nn$ measurements of cross-correlations over two-point cross-correlations in an idealized scenario. Therefore, the absolute numbers presented in Fig. \ref{fig:cross_corr} and Table \ref{tab:part_constraints} should be interpreted accordingly. We have used the simulation volume of $(1\hgpc)^3$ throughout our analysis - current cosmological surveys have orders of magnitude larger volumes, and the absolute constraints are expected to be even tighter, modulo survey systematics. Second, we have only used measurements for the first four nearest neighbors in this analysis. As pointed out in Sec. \ref{sec:measurement}, extending to higher $k$ does not add any significant computational cost. Including these higher $k$ measurements can further improve the constraints from the $k\nn$ cross-correlations, as shown in \citet{Banerjee_Abel}. However, it should be kept in mind that increasing the length of the data vector, at fixed number of realizations, reduces the Hartlap factor, and the analysis might be unstable if the data vector is too large. Third, we have only used measurements of joint $k\nn$-$\cdf$s when $k_1=k_2$. It is also possible to consider cases where $k_1\neq k_2$, again, without significant additional costs. Since each combination has a unique expression in terms of various correlation functions (see Sec. \ref{sec:theory}), adding them to the analysis can also help improve sensitivity to various cosmological parameters. The limitation, once again, would be ensuring that the data vector size does not become comparable to the number of realizations used to estimate the covariance matrix.

\subsection{Detecting cross-correlations in sparse samples}
\label{sec:sparse samples}

As a final example of the use of joint $k\nn$-$\cdf$ for describing spatial correlations of different datasets, we consider the problem of detecting spatial correlations in sparse samples. Under these conditions, measurement noise from a finite number of tracers can dominate the signal even if a true correlation exists in the underlying continuous fields.

As a concrete example, we choose $1000$ halos at random from the most massive $10^5$ halos in one of the realizations from the fiducial cosmology of the \quijote  simulations. Next, we choose another $1000$ halos at random (without replacement from the ones chosen in the first set) from the same box. These two sets of halos should be correlated spatially since they both trace the same underlying field. We then compute the two-point correlation function between these two sets, $\xi_d(r)$ using \corrfunc, as well as by computing the joint first nearest neighbor distribution $\cdf_{1,1}(r)$ over the same set of scales --- $10 \hmpc$ to $100\hmpc$, using $40$ measurement bins. Once again, since we are looking to quantify the cross-correlation between the sets, we remove the signal from the clustering of the individual datasets by using Eq. \ref{eq:xi_tilde}, and using only $\psi^{(1,1)}_d (r)$ in the rest of our analysis. For the nearest neighbor measurements, we use $4\times 10^6$ random points distributed over the simulation volume.

Next, keeping the first sample of $1000$ halos fixed, we iterate over $1000$ other realizations of the fiducial cosmology of the \quijote simulations. For each realization, we choose $1000$ halos at random from the $10^5$ most massive halos in the box. Note that since the cosmology is held fixed here, the amplitude of clustering in all of these halo samples should roughly be the same. We then compute the spatial correlations -- in terms of both the two-point cross correlation, and the joint $k\nn$ distributions -- of these $1000$ samples from different realizations with the original sample of halos. For halo samples drawn from different realizations of the same cosmology, no correlation is expected, since they sample different modes, and therefore the spatial fluctuations of counts are not correlated with each other.  

Using the measurements of cross-correlations from the $1000$ uncorrelated samples, we find the mean signal, along with the full covariance matrix. First, we consider the two-point correlation function measurements. Let us denote the mean data vector of the measurements of uncorrelated samples by $\xi_0(r)$, and the covariance matrix by $\mathbf C$. For sufficiently large number of realizations, we should have $\xi_0(r) \to 0$. The values of $\xi(r)-\xi_0(r)$ for $50$ of the the realizations are plotted using the (light) solid lines on the left panel of Fig. \ref{fig:ratio}. Notice that the measured signal is not $0$, even though the samples should be uncorrelated - this is due to noise introduced by the sparsity of the samples. Next, the $\chi^2$ value can be found for each of the $1000$ realizations by computing
\eq{chi_sq}{\chi^2_i = \Big(\xi_i(r) - \xi_0(r)\Big)^T\mathbf C^{-1}\Big(\xi_i(r) - \xi_0(r)\Big)\, ,}
where the label $i$ denotes any one of the $1000$ realizations. The distribution of $\chi^2$ over all the realizations is represented by the lighter colored histogram in Fig. \ref{fig:chi2}. We use measurements from $40$ bins, and so nominally have $\sim 40$ degrees of freedom, as some of the measurements will be correlated. This is consistent with the fact that the $\chi^2$ distribution peaks near $40$. 

For the case where there is an underlying correlation between the halo samples, i.e. when the two halo  samples are drawn from the same realization, we plot $\xi_d(r) - \xi_0(r)$ using the dashed (dark) line in the left panel of Fig. \ref{fig:ratio}. We also compute the value of $\chi^2$ in this instance by replacing $\xi_i(r)$ by $\xi_d(r)$ in Eq. \ref{eq:chi_sq}. This value is represented by the dotted vertical line in Fig. \ref{fig:chi2}. Notice that this value lies well within the range of $\chi^2$ obtained for samples drawn from the null hypothesis - \textit{i.e.} uncorrelated samples. This is mainly due to the fact that the samples are sparse, and therefore, the measurement noise dominates over the true signal. This latter fact can also be seen by comparing the dashed and solid lines in the left panel of Fig. \ref{fig:ratio}.

We repeat this procedure outlined above for the joint $\nn$ measurements. We denote the mean of the measurement for uncorrelated samples by $\psi^{(1,1)}_0(r)$ and the covariance matrix by $\mathbf {\tilde C}$. The values of $\psi^{(1,1)}(r)-\psi^{(1,1)}_0(r)$ for $50$ of the uncorrelated samples are plotted using the (lighter) solid lines on the right panel of Fig. \ref{fig:ratio}. The $\chi^2$ values are obtained in this case by simply replacing $\xi(r)$ by $\psi^{(1,1)}(r)$, and $\mathbf C$ by $\mathbf {\tilde C}$ in Eq. \ref{eq:chi_sq}. The distribution of $\chi^2$ for uncorrelated samples as measured through the joint nearest neighbors is represented by the darker histogram in Fig. \ref{fig:chi2}. As with the two-point correlation function, $40$ bins were used, and the distribution peaks near $40$, roughly $1$ per degree of freedom. Comparing the two histograms, we see that the two-point correlation function and the nearest neighbor measurements perform the same for uncorrelated samples - \textit{i.e.} in the absence of a true clustering signal. Next, we consider the nearest neighbor measurements for the correlated sample --- where the halos were drawn from the same realization. We plot  $\psi^{(1,1)}_d(r) - \psi^{(1,1)}_0(r)$ with the dashed (darker) line on the right panel of Fig. \ref{fig:ratio}. For the nearest neighbor measurements, it is clear that the measurement for the correlated sample is clearly separated from the measurements on the uncorrelated samples. To quantify this separation, we compute the $\chi^2$ value for the correlated samples, using the measurement of $\psi^{(1,1)}_d(r)$. This value is represented by the dashed vertical line in Fig. \ref{fig:chi2}, and, unlike in the two-point correlation case, is far to the right of the distribution of $\chi^2$ values for the uncorrelated samples. Since we used $1000$ different uncorrelated halo samples to characterize the $\chi^2$ distribution, and none of these produce as large a $\chi^2$ value, we can summarize this difference in terms of the $p$-value: $p<10^{-3}$. Note that the shape of the distribution of the $\chi^2$ values, and the actual value of $\chi^2$ obtained for the correlated sample, suggest that actual $p$-value is likely much smaller. Therefore, the  hypothesis that the two halo samples from the same underling distribution are uncorrelated is highly disfavored when the correlation is measured through the nearest neighbor method.

From this simple example, we can conclude that spatial correlations between sparse datasets can be detected more robustly when using the joint $\nn$ distributions compared to the two-point correlation functions. The fact that the $\chi^2$ distributions for the uncorrelated samples, as measured through either $\xi(r)$ or $\psi^{(1,1)}(r)$, match closely 
indicates that the joint $\nn$ distribution is truly detecting the underlying cross correlation, rather than being a numerical artifact. We note once again that the same input data, \textit{i.e.} the same sets of halo positions, is used in both calculations --- the difference lies only in the way the data is summarized. It is also worth noting that we have only used $\cdf_{1,1}$ to capture the correlations. In principle, we could have used other joint $k\nn$ measurements as well, adding to the statistical significance of the detection. Since our purpose here is only to demonstrate the usefulness of the joint $\nn$ correlation measure, and given that the $\cdf_{1,1}$ is already sufficient to demonstrate this, we do not delve further into combining the higher $k\nn$ measurements in this section. For real datasets, the choice of which joint $k\nn$ distributions to use will depend on the goals of the analysis, as well as the features of the datasets under consideration.


\section{Summary and Discussion}
\label{sec:conclusions}

In this paper, we have extended the nearest neighbor framework for measuring clustering developed in \citet{Banerjee_Abel} to include the joint nearest neighbor distributions of two different datasets and use these to measure cross-correlations in the spatial clustering of the two. We have demonstrated that this way of measuring cross-correlations is generally more powerful than through two-point cross-correlations. Since cross-correlations between datasets are used widely and in various different contexts in cosmology, this extension greatly increases the range of analyses that the nearest neighbor measurements can be applied to. We now summarize the main points of this work.

We have shown that \textit{joint} $k\nn$-$\cdf$s are sensitive to all $N$-point correlation functions that can be formed from the two fields whose tracers we consider. Therefore, these $k\nn$-$\cdf$ can capture correlations beyond just the linear correlations between datasets - the latter is what is generally measured through the two-point cross correlations. We have outlined how the relevant distributions can be computed from data. This is done by considering the distances from random points in the volume to the $k_1$-th nearest neighbor from the first set, and the $k_2$-th nearest neighbor from the second set, and then considering the distribution of the larger of these two distances. The cross-correlation piece can be isolated by subtracting the product of the individual $k\nn$ (for each dataset) distributions from the joint $k\nn$ distribution. We reiterate that this entire calculation is really fast and can be performed in less than a minute on a single core for typical applications explored in this paper.

We have demonstrated the applications of these measurements, first in the context of Gaussian fields, where it is possible to analytically predict these distributions for both correlated and uncorrelated tracers, and then in the context of fully nonlinear cosmological fields. In the latter context, one can still predict the joint $k\nn$ distribution for uncorrelated samples in terms of the measurement of the $k\nn$ measurements on each sample separately. Using a Fisher matrix formalism, we have shown that cross-correlations of massive halos with the underlying matter field as measured through $k\nn$ measurements are more sensitive to the underlying cosmological parameters, compared to measurements of the two-point cross-correlation. Finally, we have demonstrated that the nearest neighbor measurements can robustly detect cross-correlations in low number-density samples where the two-point cross-correlation measurements are dominated by measurement noise. For both of these applications, we use only those parts of the $k\nn$ measurements that come only from the cross-correlations of the datasets, to enable a direct comparison to the two-point cross-correlation function. Unlike the two-point cross-correlation, a lack of detection of cross-correlations using the $k\nn$ distributions imply a complete statistical independence of the two distributions from which the tracers are drawn. 

 It is worth noting that the specific method outlined here is not the only nearest neighbor measurement approach that is sensitive to cross-correlations of two datasets. For example, combining the two sets of tracers, and performing $k\nn$ measurements on this combined set should also, in principle, be sensitive to the cross-correlations of the two sets. Another possible way to measure cross-correlations in this framework is to compute distance to a specific $k$-th nearest neighbor data point from the first data set, and then computing the distribution of how many data points from the second set are found within that volume. However, there are certain attractive features of the method presented here: first, the expression for the joint $k\nn$-$\cdf$s can be conveniently expressed in terms of the various $N$-point correlation functions formed from the two fields. Second, it is both conceptually and computationally easy to isolate those parts of the measurement which depends on just the cross-correlations. This factorization is useful for various common applications in cosmology, and is not guaranteed for other ways of measuring cross correlations with nearest neighbors.

Another aspect worth noting is that the joint $k\nn$ measurements outlined here are applied to cross-correlations between two samples. It is possible to extend this formalism to include more datasets. the simplest way to consider various pairs of datasets and consider their cross-correlations - this is routinely done using the two-point cross-correlations. However, it is also possible to consider multiple sets of tracers at the same time, and consider whether they are all drawn from the same distribution.

Lastly, we have focused on cross-correlations between discrete datasets in this paper. However, as illustrated in Sec. \ref{sec:Gaussian}, the $k\nn$ measurements on discrete tracers are able to capture the cross-correlations of the underlying continuous fields, even at the level of downsampling we have used - the $k\nn$ measurements are performed on only $2\times 10^5$ particles out of the $512^3$ particles that describe the density field in the simulations. This is similar to the findings for the $k\nn$ auto-correlation measurements in \citet{Banerjee_Abel}. Taken together, this implies that the formalism presented here can be extended to measuring cross-correlations between continuous maps, by simply sampling the maps correctly with a fixed number of tracers. We will explore this aspect in detail in future, to widen the range of applications of $k\nn$-$\cdf$ cross-correlations to datasets which are inherently continuous.

\section*{Acknowledgements}

This work was supported by the Fermi Research Alliance, LLC under Contract No. DE-AC02-07CH11359 with
the U.S. Department of Energy, and the U.S. Department of Energy SLAC
Contract No. DE-AC02-76SF00515. The authors thank Alvaro Zamora for comments and discussion which helped improve the paper. Some of the computing for this project was performed on the Sherlock cluster. The authors would like to thank Stanford University and the Stanford Research Computing Center for providing computational resources and support that contributed to these research results. The \textsc{Pylians3}\footnote{https://github.com/franciscovillaescusa/Pylians3} analysis library  was used extensively in this paper. We also acknowledge the use of the \textsc{GetDist}\footnote{https://getdist.readthedocs.io/en/latest/} \citep{2019arXiv191013970L} software for plotting.

\section*{Data Availability}

The simulation data used in this paper is publicly available at \url{https://github.com/franciscovillaescusa/Quijote-simulations}. Additional data is available on reasonable request.



\bibliographystyle{mnras}
\bibliography{ref} 




\appendix


\section{Derivation of the generating function for joint counts}
\label{sec:derivation}

Here we extend the formalism from \citet{Szapudi1993} and Appendix A in \citet{Banerjee_Abel} to two continuous fields, instead of one. For two continuous fields $\rho_1(\boldsymbol r)$ and $\rho_2(\boldsymbol r)
$, the generating functional for all correlation functions can be written as an integral over all possible joint configurations of the two fields:
\eq{continuous}{\mathcal Z\left[J_1,J_2\right] = \int \big[D \rho_1\big] & \big[D \rho_2\big] \mathcal P\big(\rho_1, \rho_2\big) \nonumber \\  & \times \exp\Bigg[i \int d^3\boldsymbol r \Big(\rho_1 J_1 + \rho_2 J_2\Big)\Bigg]\, ,}
where $\mathcal P(\rho_1, \rho_2)$ represent the joint distribution function of the two fields. Note that it is this joint distribution function that encodes any correlation between the two fields - in the absence of correlations, the joint distribution can be factorized into the individual distribution functions of the two fields. The connected $N$-point correlations, $\xi^{(k_1,k_2)}$, with $k_1$ factors of $\rho_1$ and $k_2$ factors of $\rho_2$, are given by the functional derivatives of the generating functional with respect to the sources $J_1$ and $J_2$:
\eq{cumulants}{\bar \rho_1^{k_1}\bar \rho_2^{k_2}&\xi^{(k_1,k_2)}(\boldsymbol r_1,...,\boldsymbol r_{k_1};\boldsymbol r^\prime_1,...,\boldsymbol r^\prime_{k_2})\nonumber \\ &= \frac{(-i)^{\left(k_1+k_2\right)}\delta^{\left(k_1+k_2\right)}\left(\log \mathcal Z\left[J_1,J_2\right]\right)}{\delta J_1(\mbr_1)...\delta J_1(\mbr_{k_1})\delta J_2(\mbr^\prime_1)...\delta J_2(\mbr^\prime_{k_2})}\Bigg|_{J_1=0,J_2=0}\, ,}
where $\bar \rho_i$ are the mean densities of the two fields. Conversely, the generating functional can be expressed in terms of the $N$-point correlation functions as
\eq{partition}{\mathcal Z\left[J_1,J_2\right] = \exp\Bigg[\sum_{k_1=0}^\infty \sum_{k_2=0}^\infty \frac{\left(i\bar \rho_1\right)^{k_1}\left(i\bar \rho_2\right)^{k_2}}{k_1!k_2!}\int d^3\mbr_1...d^3\mbr_{k_1}\nonumber \\ \times d^3\mbr^\prime_1...d^3 \mbr^\prime_{k_2}\xi^{(k_1,k_2)}(\mbr_1,...,\mbr_{k_1};\mbr^\prime_1,...,\mbr^\prime_{k_2})\nonumber \\ \times J_1(\mbr_1)...J_1(\mbr_{k_1})J_2(\mbr^\prime_1)...J_2(\mbr^\prime_{k_2})\Bigg]\, .}
For a set of tracers of these underlying fields generated by a local poisson process, the number of tracers of each type contained in volume $V$ around a point $\mbr$ depends on the integral of the field over the same volume
\eq{volume_integral}{\mathcal M^{(i)}_V(\mbr) = \int d^3 \mbr^\prime \rho_i(\mbr^\prime)W(\mbr, \mbr^\prime)\,,}
where $W$ represents the window function for smoothing the fields.

The probability of finding $k_i$ tracers of type $i$, in a sphere of volume $V$, is given by
\eq{poisson_sampling}{P\left(k_i|\mathcal M_V^{(i)}\right) = \frac{\left(\mathcal M_V^{(i)}/m_i\right)^{k_i}}{k_i!}\exp\Bigg[-\frac{\mathcal M_V^{(i)}}{m_i}\Bigg]\,,}
where $m_i$ is the ``mass'' associated with tracer of type $i$. The \textit{joint} probability of finding $k_1$ tracers of field $\rho_1(\mbr)$ and $k_2$ tracers of field $\rho_2(\mbr)$ in a volume $V$ can be written in terms of the integral $\mathcal M_V^{(1)}$ and $\mathcal M_V^{(2)}$ as
\eq{joint_sampling}{P\left(k_1,k_2|\mathcal M_V^{(1)}, \mathcal M_V^{(2)}\right) =& \frac{\left(\mathcal M_V^{(1)}/m_1\right)^{k_1}}{k_1!}\exp\Bigg[-\frac{\mathcal M_V^{(1)}}{m_1}\Bigg]\nonumber \\ &\times \frac{\left(\mathcal M_V^{(2)}/m_2\right)^{k_2}}{k_2!}\exp\Bigg[-\frac{\mathcal M_V^{(2)}}{m_2}\Bigg]\,.}
The probability of $k_1$ and $k_2$ at a fixed volume $V$, or equivalently, radius $R$, can be computed by averaging over all $\mathcal M_V^{(i)}(\mbr)$:
\eq{average_joint}{P\left(k_1,k_2|V\right) =  \Bigg\langle & \frac{\left(\mathcal M_V^{(1)}/m_1\right)^{k_1}}{k_1!}\exp\Bigg[-\frac{\mathcal M_V^{(1)}}{m_1}\Bigg]\nonumber \\ & \times \frac{\left(\mathcal M_V^{(2)}/m_2\right)^{k_2}}{k_2!}\exp\Bigg[-\frac{\mathcal M_V^{(2)}}{m_2}\Bigg]\Bigg\rangle \, .}
Note that if the two fields $\rho_1$ and $\rho_2$ are correlated, then fluctuations in $\mathcal M_V^{(i)}$ are correlated, and the above average does not factor into the individual averages over $\mathcal M_V^{(1)}$ and $\mathcal M_V^{(2)}$. 

The generating function for the discrete counts can be written as 
\eq{discrete_generating_function}{P\left(z_1,z_2|V\right) &= \sum_{k_1=0}^\infty\sum_{k_2=0}^\infty P \left(k_1,k_2|V\right) z_1^{k_1}z_2^{k_2}\nonumber \\ &= \Bigg\langle \exp\Bigg[\frac{\mathcal M_V^{(1)}}{m_1}\left(z_1-1\right) + \frac{\mathcal M_V^{(2)}}{m_2}\left(z_2 - 1\right)\Bigg]\Bigg\rangle \, .}
While Eq. \ref{eq:discrete_generating_function} is the generating function for the joint discrete counts, the average over $\mathcal M_V^{(i)}$ can be evaluated in terms of the underlying continuous fields:
\eq{discrete_continuous_connection}{P\left(z_1, z_2 | V\right)= \int &  \left[D \rho_1\right]\left[D \rho_2\right]\mathcal P\left(\rho_1, \rho_2\right) \nonumber \\ & \times \exp\Bigg[\frac{\left(z_1-1\right)}{m_1}\int_V d^3\mbr^\prime \rho_1(\mbr^\prime)W(\mbr, \mbr^\prime)\nonumber \\ & \qquad \quad + \frac{\left(z_2-1\right)}{m_2}\int_V d^3\mbr^\prime \rho_2(\mbr^\prime)W(\mbr, \mbr^\prime)\Bigg]\, .}
With the following identification,
\eq{current_substitution}{J_i(\mbr^\prime) = W(\mbr, \mbr^\prime) \frac{\left(z_i-1\right)}{im_i}\,,}
Eqs. \ref{eq:discrete_continuous_connection} and \ref{eq:continuous} are equivalent. Then, noting that the RHS of Eqs \ref{eq:continuous} and \ref{eq:partition} both define the same quantity, and focusing on the spherical top-hat smoothing window for $W(\mbr, \mbr^\prime)$, Eq. \ref{eq:discrete_continuous_connection} can be written as 
\eq{final_expression}{P\left(z_1,z_2|V\right) = \exp\Bigg[&\sum_{k_1=0}^\infty \sum_{k_2=0}^\infty \frac{\bar n_1^{k_1}(z_1-1)^{k_1}}{k_1!}\frac{\bar n_2^{k_2}(z_2-1)^{k_2}}{k_2!}\nonumber \\ &\times \int_V d^3\mbr_1...d^3\mbr_{k_1}d^2\mbr_1^\prime ...d^3\mbr_{k_2}^\prime \xi^{(k_1,k_2)}\Bigg]\, , }
where $\bar n_i = \bar \rho_i/m_i$ represents the mean number density of each sample of tracers. Eq. \ref{eq:final_expression} makes explicit the fact that the generating function of the joint counts is sensitive to all possible correlations functions that can be constructed from the two underlying fields. In the absence of statistical correlations, the generating function for the discrete counts factorizes into two independent generating functions, one for each set of tracers.

\section{Joint cumulative counts for correlated Gaussian fields}
\label{sec:Gaussian_expressions}

In this section, we present the analytic expressions for some of the joint $k\nn$-$\cdf$s for a Gaussian random field. In particular, we look at the cases $k_1=k_2=1$ and $k_1=k_2=2$, \textit{i.e.} the first and second joint nearest neighbor distributions. For $k_1=k_2=1$, 
\eq{C_k1}{{\rm CDF}_{1,1}(r) = \mathcal P(>\ns 0, >\ns 0|V) = C(z_1,z_2|V)\Big |_{z_1,z_2=0}\,,}
where $C(z_1,z_2|V)$ is given by Eq. \ref{eq:cumulative_generating_function}. For the Gaussian case, we use Eqs. \ref{eq:Gaussian_generating_function} and \ref{eq:gaussian_single}, and therefore,
\eq{gaussian11}{\mathcal P(>\ns 0, >\ns 0|V) = 1 &- \exp\Bigg[-\bar n_1 V + \frac 1 2 \bar n_1^2 \bar \xi^{(2,0)}_{V}\Bigg] \nonumber \\ &- \exp\Bigg[-\bar n_2 V + \frac 1 2 \bar n_2^2 \bar \xi^{(0,2)}_{V}\Bigg] \nonumber \\ &+ \exp\Bigg[-\bar n_1V -\bar n_2 V+\frac 1 2 \bar n_1^2 \bar \xi_V^{(2,0)} \nonumber \\ & \qquad \quad + \frac 1 2 \bar n_2^2 \bar \xi_V^{(0,2)} + \bar n_1 \bar n_2 \xi_V^{(1,1)}\Bigg]\,.}
For the joint second nearest neighbor distribution, we use the fact that 
\eq{gaussian22}{{\cdf_{2,2}(r) = \mathcal P(>\ns 1, >\ns 1|V) = \frac{{\rm d}^2C(z_1,z_2|V)}{{\rm d}z_1{\rm d}z_2}}\Bigg|_{z_1,z_2 = 0}\,.}
After performing the derivatives with respect to $z_1$ and $z_2$, the joint CDF can be expressed as 
\eq{gaussian22a}{ \mathcal P(>\ns 1, >\ns 1|V)=& 1 - \mathcal P_1(0|V) - \mathcal P_2(0|V) - \mathcal P_1(1|V) - \mathcal P_2(1|V) \nonumber \\ &+ \mathcal P(0,0|V) + \mathcal P(1,0|V)+\mathcal P(0,1|V)\nonumber \\ & +\mathcal P(1,1|V)\,,}
with 
\eq{gassian0}{ \mathcal P_1(0|V) = \exp\Bigg[-\bar n_1 V + \frac 1 2 \bar n_1^2 \bar \xi^{(2,0)}_{V}\Bigg] \, ,}
\eq{gassian0a}{\mathcal P_2(0|V) = \exp\Bigg[-\bar n_2 V + \frac 1 2 \bar n_2^2 \bar \xi^{(0,2)}_{V}\Bigg] \, ,}
\eq{gassian1}{\mathcal P_1(1|V) = \Bigg(\bar n_1 V + \bar n_1^2 \bar \xi^{(2,0)}_{V} \Bigg)\exp\Bigg[-\bar n_1 V + \frac 1 2 \bar n_1^2 \bar \xi^{(2,0)}_{V}\Bigg] \, ,}
\eq{gassian1a}{\mathcal P_2(1|V) = \Bigg(\bar n_2 V + \bar n_2^2 \bar \xi^{(0,2)}_{V} \Bigg)\exp\Bigg[-\bar n_2 V + \frac 1 2 \bar n_2^2 \bar \xi^{(0,2)}_{V}\Bigg] \, ,}
\eq{Gaussian_00}{\mathcal P(0,0|V) = \exp \Bigg[-\bar n_1V - \bar n_2 V +\frac 1 2 \bar n_1^2\bar \xi^{(2,0)}_V + \frac 1 2 \bar n_2^2 \bar \xi^{(0,2)}_V \nonumber \\
 + \bar n_1 \bar n_2 \bar \xi^{(1,1)}_V \Bigg]\, ,}
\eq{Gaussian_10}{\mathcal P(1,0|V) = \mathcal P(0,0|V)\times \Big(\bar n_1 V - \bar n_1^2 \bar \xi^{(2,0)}_V - \bar n_1 \bar n_2 \bar \xi^{(1,1)}_V\Big)\,,}
\eq{Gaussian_01}{\mathcal P(0,1|V) = \mathcal P(0,0|V)\times \Big(\bar n_2 V - \bar n_2^2 \bar \xi^{(0,2)}_V - \bar n_1 \bar n_2 \bar \xi^{(1,1)}_V\Big)\,,}
\eq{Gaussian_11}{& \mathcal P(1,1|V)  = \mathcal P(0,0|V)\times\Bigg(\bar n_1 \bar n_2 \bar \xi^{(1,1)} + \nonumber \\ &\Big( \bar n_1 V - \bar n_1^2 \bar \xi^{(2,0)}_V - \bar n_1 \bar n_2 \bar \xi^{(1,1)}_V
\Big) \Big(\bar n_2 V - \bar n_2^2 \bar \xi^{(0,2)}_V - \bar n_1 \bar n_2 \bar \xi^{(1,1)}_V\Big) \Bigg)\, .}
Since $\bar \xi^{(2,0)}_V=V^2\sigma_1^2(r)$, and $\bar \xi^{(0,2)} = V^2 \sigma_2^2(r)$, where $\sigma^2(r)$ is the variance of fluctuations on scale $r$, these expressions above can be directly evaluated if the variance for each field, along with the degree of cross-correlation is known. For the specific case when both sets of tracers under consideration trace the same matter field, $\bar \xi^{(1,1)}_V = V^2\sigma^2(r)$, and so all the expressions can be evaluated with the help of, for example, the \textsc{Colossus} software, which returns the linear theory value for the variance as a function of scale and redshift. In the case where there are no cross-correlations, we set $\bar \xi^{(1,1)}_V=0$ for all $V$, and evaluate the above expressions accordingly.

\bsp	
\label{lastpage}
\end{document}